# The creep deformation of a new nickel-base alloy-by-design studied using synchrotron X-ray diffraction

Jingwei Chen [a], Zifan Wang [a], Chrysanthi Papadaki [b], Alexander M. Korsunsky [a] *

a: MBLEM, Department of Engineering Science, University of Oxford, Parks Road, Oxford OX1 3PJ, United Kingdom

b: Williams Advanced Engineering, Wantage, Oxfordshire, OX12 0DQ, United Kingdom

jingwei.chen@eng.ox.ac.uk

zifan.wang@exeter.ox.ac.uk

chrysanthi.papadaki@wae.com

Alexander.korsunsky@eng.ox.ac.uk, *corresponding author

## Abstract

Understanding the creep mechanisms and deformation response at different stresses and temperatures is crucial for design using nickel-base superalloys for high-temperature applications. In this study, the creep behaviour of a newly designed superalloy (nominated Alloy 11) at 750 °C/400MPa and 700°C/800MPa was systematically investigated using SEM, STEM, EBSD and synchrotron X-ray diffraction. Material properties and mechanical response were analysed by considering such properties as lattice parameters and misfit, phase volume fraction, microstrain in reference and crept samples. Detwinning and dynamic recrystallization were observed for the crept samples. STEM characterization of deformed materials reveals that multiple deformation mechanisms and defects could be identified in crept samples, namely, dislocations in the $\gamma$ matrix channels and in $\gamma'$ precipitates, along with continuous and isolated stacking faults. The highest values of lattice misfit, microstrain and lattice parameter change were observed at the centre of dogbone-shaped crept samples. This was associated with the hump-shaped temperature distribution profile during creep testing. The significant spatial variation detected in terms of lattice parameters, misfit and microstrain support the conclusion that it is important to perform spatially resolved measurements instead of considering each sample as a single measurement point when investigating creep response. It is also worth noting that the statistics of the lattice-parameter distribution for the $\gamma$ and $\gamma'$ phases obey a Gaussian distribution. In conclusion, a discussion is given for the meaning and implications of these findings for future research.

## 1. Introduction

Nickel-base superalloys exhibit high strength and creep resistance at elevated temperatures. These properties, together with their fatigue performance, microstructural stability, and high-temperature corrosion resistance have made them the materials of choice for the hot section of gas turbines, such as blades, discs, vanes, and combustion casings.

The need for greater efficiency of engines causes an increase in the turbine inlet temperature and thus requires the development of stronger, microstructurally and compositionally more complex, and more creep-resistant superalloys [1]. The temperature profile and stress states during a typical flight cycle loading are complex and non-proportional so both high-stress/high-temperature and high-stress/low-temperature combinations are encountered, causing detrimental creep deformation of nickel superalloys.

The peculiar thermo-mechanical performance of nickel-base superalloys mainly arises from the presence of an intermetallic phase, $\gamma'$ precipitates $Ni_3(Ti, Al)$, which have $L1_2$ structure (an ordered FCC arrangement). When dislocations travel through the matrix and encounter $\gamma'$ particles, they are prevented from easy glide and must undergo bowing, climb, splitting, and formation of anti-phase boundaries (APB) for further deformation to occur. Thus, the deformation resistance of an alloy is strongly dependent on the volume fraction of $\gamma'$ precipitates that can be modified by the thermo-mechanical treatment history, i.e. the combination of temperature and applied load that has been shown to have a tremendous effect on the creep deformation [1]. A modern generation of superalloys is characterised by a high-volume fraction $\gamma'$ to improve the resistance to plastic deformation (creep).

Creep is the phenomenon of time-dependent deformation of a material when subjected to constant load or stress. A better understanding of creep damage mechanisms improves our understanding of the internal stress and strain interactions during the thermo-mechanical loading of polycrystalline alloys. The ultimate purpose of the analysis is to underpin the development of improved material design, including through the optimization of thermomechanical processing routes. A considerable amount of research work has been conducted on the creep deformation in nickel superalloys over the past 50 years. Many creep deformation mechanisms were proposed for different stress and temperature regimes, such as isolated and continuous stacking faults [3], [4], [5], anti-phase boundary (APB) shearing [3], [6], [7], [8], [9], dislocation climb [10], [11], [12], [13] and microtwinning [14], [15], [16], [17], [18]. These mechanisms may operate simultaneously during creep at a given specific combination of stress and temperature. Diffusion creep is a mechanism involving the diffusion of atoms or material transport. At low-stress and high-temperature loading conditions, diffusional and time-dependent dislocation climbs are commonly favourable deformation mechanisms. There are mainly two types of diffusion creep, depending on whether diffusion occurs through grain boundaries, termed Coble creep (activated at a lower temperature) or through the main body of the grains, known as Nabarro-Herring creep (favoured at high temperatures). Dislocation creep occurs by the motion of dislocations through crystal lattices. Dislocations can glide along slip planes, and this requires little thermal activation, hence, dislocation creep becomes favourable at relatively low temperatures. The dominant creep deformation mechanisms at high stress and low temperature are APB shearing and stacking fault [3,8].

Transmission X-ray diffraction (XRD) is one of the most powerful methods for residual strain and stress analysis that can achieve sub-micron resolution, particularly with the help of synchrotron X-ray beams. Particle diffraction-based methods (including neutrons here) [19, 20, 21, 22] evaluate the residual stresses by determining the changes in interplanar lattice spacing. Bragg's law provides a concise mathematical description of the relationship between lattice spacing, scattering angle and incident beam wavelength. In the monochromatic mode, the scattering angle is measured precisely, so that for known incident beam wavelength the changes in lattice spacing can be found using Bragg's law. Figure 1 gives an example of a typical diffraction pattern, which is presented as detector counts versus scattering angles. It is illustrated that peak centre position and shape (height, half-width change, integrated intensity), quantitative information can be extracted regarding lattice strain/compositional changes, atomic density and permanent deformation. Synchrotron XRD strain mapping has been widely employed to evaluate the time-independent strain at locations of particular interest, such as near crack tips [23], [24], [25]. However, there appears a dearth of information on the spatially resolved strain distributions after time-dependent creep deformation using this technique. The main advantage of synchrotron XRD is that it can simultaneously and accurately provide a quantitative description of lattice parameters, phase fraction and microstrain in dual-phase polycrystalline materials.

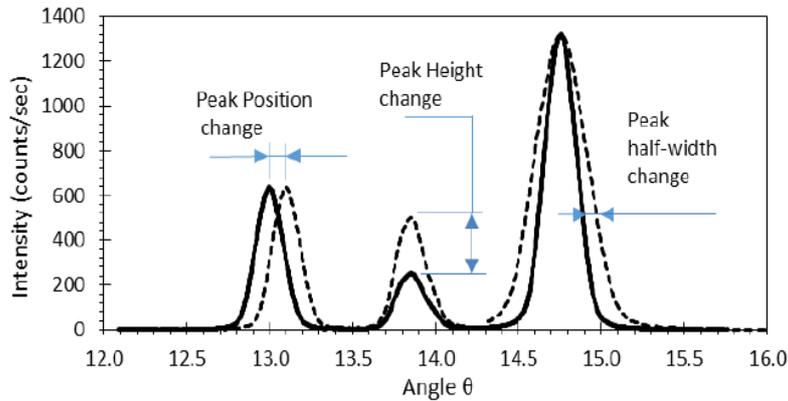

**Figure 1**. A typical diffraction pattern [26].

In the present investigation, two dog-bone-shaped samples were pre-crept at different temperature/stress combinations and stopped within the secondary creep regime. Ex-situ synchrotron was then performed on one reference sample and two crept samples. The distribution of lattice-parameter changes in $\gamma$ and $\gamma'$ phase was studied. The current research is aimed at investigating the lattice misfit, phase fraction, lattice parameter and microstrain mapping in pre-crept Alloy 11 samples using synchrotron X-ray diffraction.

## 2. Experimental procedure

### 2.1 Materials and creep tests

The materials investigated in this study were obtained from a pancake forging of a newly developed nickel-base superalloy-by-design referred to as Alloy 11 [21, 27]. The forging part was manufactured by isothermally forging a powder isostatic pressed cylindrical ingot. Its nominal chemical composition is given in Table 1. Alloy 11 is a newly developed powder metallurgy superalloy that are particularly suitable for producing gas turbine rotor discs at higher operating temperature (generally between 700°C and 800°C). It is also relatively lightweight, resistant to environmental degradation, inexpensive and suitable for low-cost manufacturing techniques. In this paper, the creep mechanism for alloy 11 between 700°C and 800°C was examined for the first time. The material was first given a solution treatment at 1180°C for 2h and then cooled to room temperature in air. Subsequently, it received a two-step ageing treatment at 850°C for 2h and air cooled to ambient temperature, followed by ageing at 800°C for 4h before cooling by air. All the cooling rate was monitored and controlled at 0.7°C/s.

**Table 1**. Chemical composition of Alloy 11 (wt.%)

| Ni  | Co    | Cr    | Ta   | W    | Al   | Ti   | Mo   | Nb   | Fe   | Mn   | Si   |
|-----|-------|-------|------|------|------|------|------|------|------|------|------|
| Bal | 15.06 | 12.69 | 4.77 | 3.22 | 3.16 | 2.84 | 2.14 | 1.44 | 0.95 | 0.48 | 0.47 |

Three dog-bone specimens (1.5 mm wide, 1.2 mm thick and 10 mm gauge length) were machined by an electro-discharge machine from the forging part. The surface of the gauge length was carefully polished to avoid any premature failure during thermomechanical loading. Two specimens were crept at 750°C/400MPa and 700°C/800MPa, respectively, by the Electro-Thermo-Mechanical Testing (ETMT) system (Instron Ltd., High Wycombe, UK) in an argon atmosphere. The names and the creep conditions for the three samples are listed in Table 2. The creep tests were interrupted within the secondary creep regime when the total creep strain reached 0.6%, and the specimens were then cooled to room temperature after removing the constant load. As shown in the creep curves of the two tests in Figure 2, it took 35h and 0.26h to achieve a creep strain of 0.6% for 750°C/400MPa and 700°C/800MPa loading conditions, respectively. The creep rate is much higher in the creep test with a larger constant load.

In the ETMT system, samples were heated by passing an electric current. The temperature was controlled by a proportional-integral-derivative controller (PID) and monitored with a thermocouple spot-welded to the centre of the specimen. A symmetrical bell-shaped temperature profile arises, with the temperature in the middle of the sample higher than at the two ends (see Figure B in Appendix A). It was caused by the fact that the ends of the sample were cooled through the relatively massive metallic grips.

**Table 2**. The name and creep conditions for the three samples in this study.

| Sample name | Creep conditions |
| --- | --- |
| Sample A | Reference sample, no creep |
| Sample B | 750°C/400MPa |
| Sample C | 700°C/800MPa |

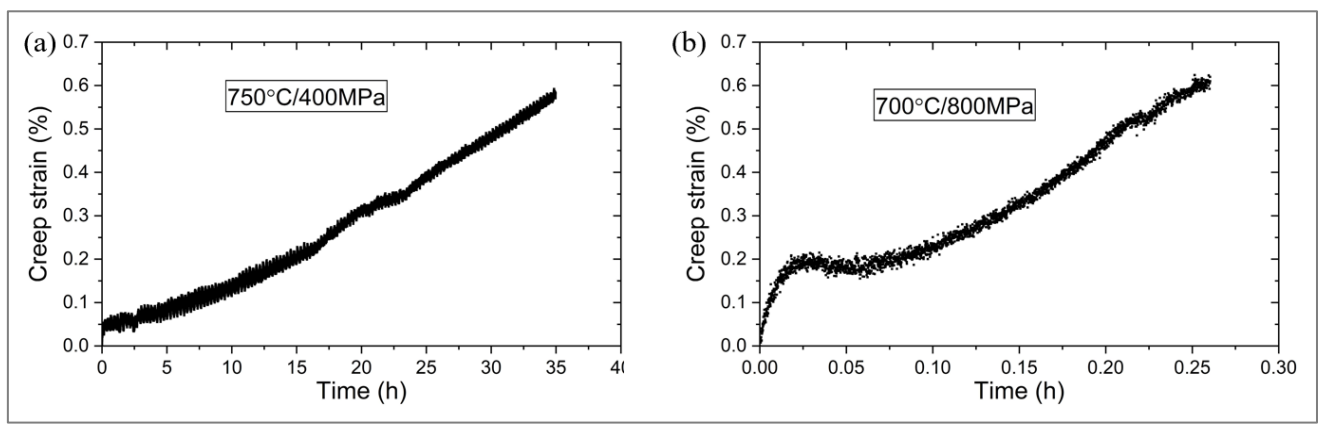

**Figure 2**. The tensile creep curves of Alloy 11 at 750°C/400MPa and 700°C/800MPa loading conditions.

## 2.2 Synchrotron XRD experiment

Synchrotron XRD experiments on one reference sample without creep and two pre-crept samples were performed at the I12 beamline at Diamond Light Source (DLS), UK. As illustrated in Figure 3(a), the dog-bone specimens were mounted horizontally with their front vertical surface perpendicular to the incident beam. Debye-Scherrer ring patterns were recorded using a large area diffraction detector Pilatus 2M (1475×1679 pixel and $172\mu m \times 172\mu m$ pixel size) that was placed behind the sample. Monochromatic beam energy of 93.7 keV was employed with a beam size fixed at $200\mu m \times 200\mu m$ and a data acquisition rate of 10s per scanning point. The position of the beam was fixed in the laboratory system, while the sample could be moved horizontally and vertically. Figure 4 illustrates the coordinate system and the scanning plan for the three samples. X axis is the horizontal direction and also the loading direction, Y axis is the vertical direction. For each sample, four scanning lines were measured in the 10mm sample gauge region, and four reference points were scanned in the grip region. Each reference measurement point was at the same height that corresponds to each scanning line. The scanning direction for each line was parallel to the loading direction. The distance between the beam centre to the sample edge and the distance between two neighbouring beam centres were $300\mu m$. $LaB_6$ powder was used for the XRD pattern calibration. Peak indexation and Rietveld refinement (as shown in figure 3(b)) were performed using the General Structure Analysis System (GSAS-II) [28]. Diffraction data from each reference point were analysed to provide the undeformed lattice parameters for each sample. To determine the phase fraction, lattice misfit, lattice parameters and microstrain (diffraction peak broadening), a series of Rietveld refinements were conducted for diffraction data from each scanning line. Following refinement, the values of residuals were of the order 5% or below, confirming the goodness and reliability of interpretation [29].

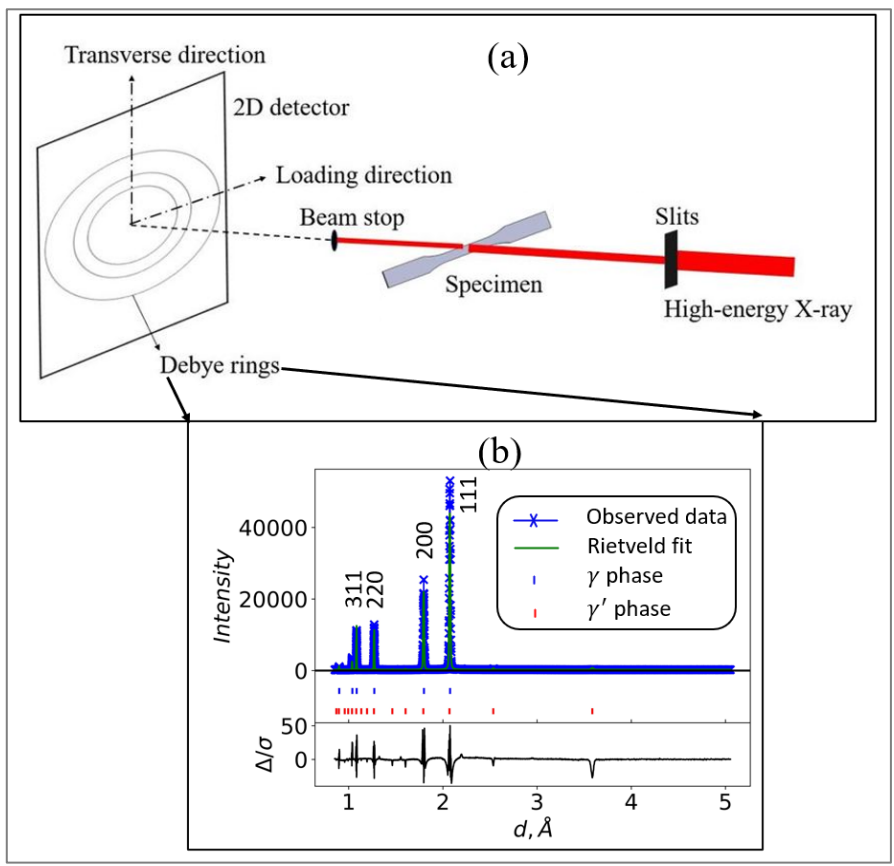

**Figure 3**. (a) Experiment setup of synchrotron X-ray diffraction at I12, DLS. (b) An example of indexed and refined diffraction peak processed by GSAS II software.

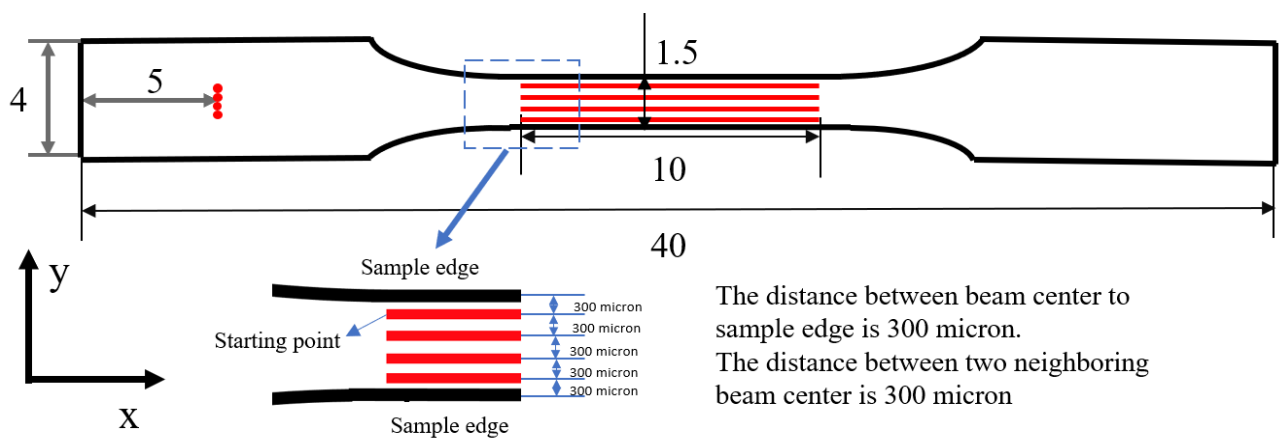

**Figure 4**. An illustration of the scanning routine for the reference sample and pre-crept samples, unit in *mm*.

## 3. Results and discussion

### 3.1 Material characterization

#### 3.1.1 EBSD characterization

After the synchrotron XRD experiment, electron backscatter diffraction (EBSD) characterization was performed using a Tescan LYRA3 FIB-SEM system equipped with a SYMMETRY CMOS EBSD detector from Oxford Instruments. The outer surface of all samples was removed by standard metallographic grinding, followed by mechanical polishing in a solution containing a suspension of diamond particles down to 1 μm. Final polishing was made with 0.25μm colloidal silica. The accelerating voltage for all EBSD mapping was 20kv and the step size was 0.5μm. Parts of the EBSD mapping along with the characterization methods were recently published in [21].

Figure 5 (a), (b-c) and (d) illustrate the EBSD measurements of Sample A, B and C, respectively. As shown in Figure 5 (a), equiaxed microstructure with twins was observed in the reference sample after a two-step ageing treatment. After thermomechanical creep deformation, no twin was found in the two crept samples. It implies that creep loading conditions at 750°C/400 MPa and 700°C/800 MPa can introduce detwinning and recrystallization in nickel-base superalloy. Figure 5 (b) and (c) depict EBSD mapping from the grip section and gauge region of the sample after creep at 750°C/400 MPa, respectively. There was no creep deformation in the grip section, while tensile and creep deformation occurred in the gauge region. The grain size in the gauge region ($9\mu m$) is much smaller than in the grip region ($27\mu m$). Close observation suggests that many of the small grains in the localized area have almost the same orientation, which means that the small grain fragments (or subgrains) come from the same grain before loading. This finding implies that significant grain rotation occurs during tensile and creep loading, and it is also consistent with previous creep experiment results in the literature [30].

It is clear in Figure 5(d) that the microstructure of Sample C at much higher stress (700°C/800MPa) is significantly different from the microstructure of Sample B at lower stress and slightly higher temperature (750°C/400MPa). Significant recrystallization occurs in Sample C and the grains are large and elongated in the loading direction during creep. Because it only takes 0.26h for Sample C compared to 35h for Sample B to get the same amount of creep deformation, a higher magnitude of stress not only dramatically increases the creep rate but also accelerates the recrystallization process. This finding is analogous to the results of dynamic recrystallization under the hot deformation or creep loading condition in face-centred cubic alloys [31-33]. Material processing parameters, such as strain rate, deformation temperature and strain level, are of significance to control the microstructural evolution during the recrystallization process. With the increase of deformation and strain rate, dislocation density within the sample becomes higher and leads to dislocation networks. Subgrains then form and eventually, grain recrystallization occurs. The apparent difference in grain morphologies for Sample B and C arises from the difference in deformation temperature and strain rate. It is assumed that Sample B was stopped at the subgrain stage while Sample C was interrupted at the recrystallization stage even though they had the same level of deformation. Additionally, the grains in the edge of Sample C are much smaller than those in the centre of the sample. It can be explained by the fact that the sample edge was cooled by the argon atmosphere. The sample centre had the highest temperature during creep, and lead to full recrystallization, as opposed to the edge of the sample.

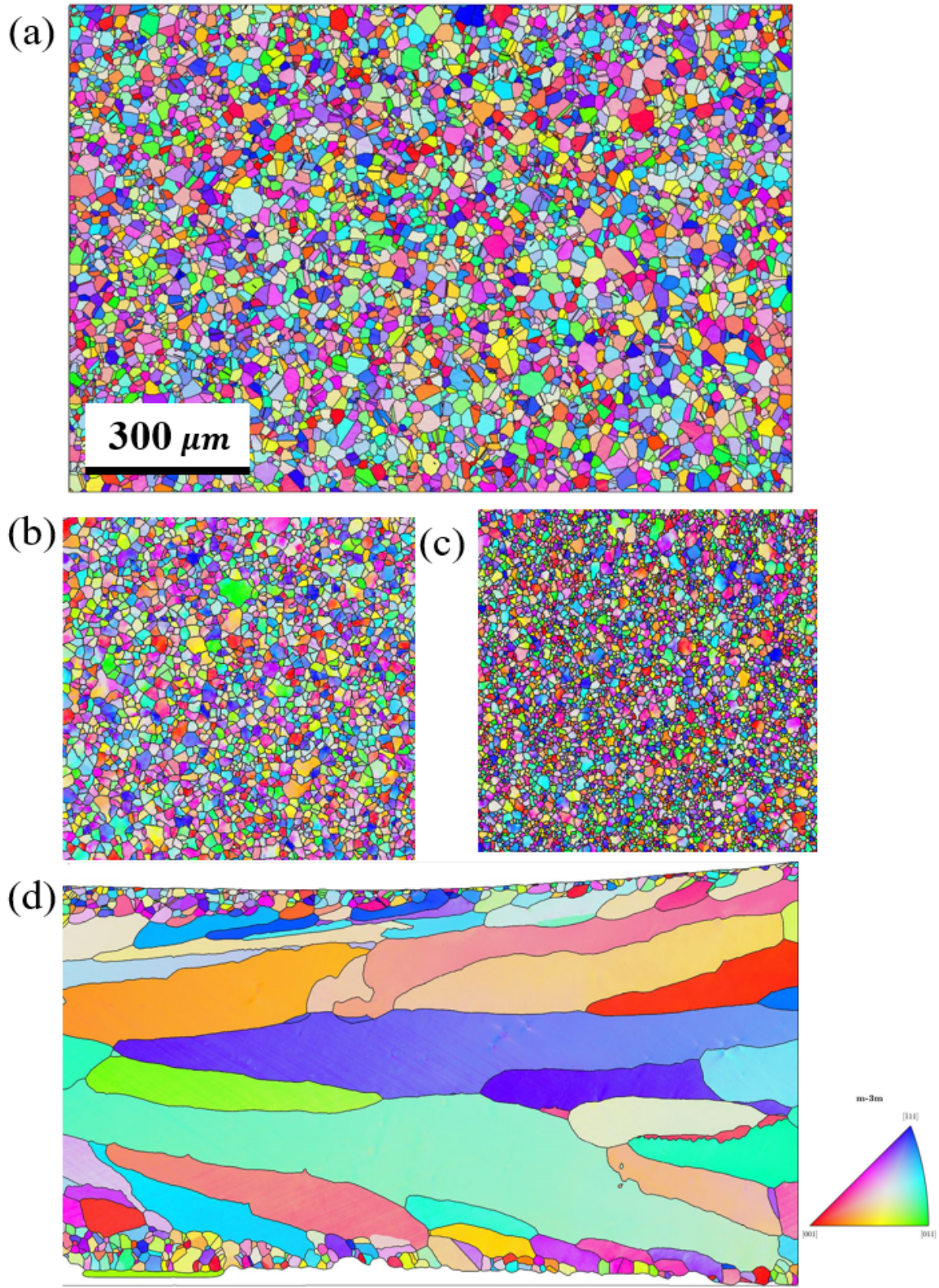

**Figure 5**. EBSD measurements of (a) reference Sample A [21], (b) the grip region of sample B crept at 750°C/400MPa [21], (c) the gauge region of Sample B crept at 750°C/400MPa (d) Sample C crept at 700°C/800MPa. The scale bar and inverse pole figure apply to all the EBSD mappings.

### 3.1.2 Deformation characterization using SEM and STEM

Scanning electron microscopy (SEM) and scanning transmission electron microscopy (STEM) characterization techniques were used to visualize the deformation mechanism of Sample B and Sample C that were crept to 0.6% plastic strain at different loading conditions. Several voids and inclusions were observed in the sample edge for both samples. Figure 6(a) reveals that both intergranular and intragranular cracks occurred on the edge of Sample B, while only intergranular cracks were found in Sample C as illustrated in Figure 6(c). Intergranular crack is a typical damage mode under creep loading, here we demonstrate that intragranular crack could also appear when there is a void in the crack path. Figure 6(b) and (d) show enlarged images of red rectangular areas A and B, respectively. Microvoids in the grain boundary and triple points were observed in Sample B, as shown in Figure 6(b). A close observation of the crack tip in sample C suggests that the crack stopped at the border of a primary $\gamma'$ precipitate.

STEM images in Figure 7 reveal the microstructure and the deformation mechanisms for alloy 11 after creeping at 700°C/800MPa. Four kinds of deformation mechanisms were observed in the current study: (1) Dislocations in the $\gamma$ matrix channel. (2) Shearing of $\gamma'$ precipitate by APBs. (3) Continuous stacking faults cutting through both $\gamma$ matrix and $\gamma'$ precipitate. (4) isolated stacking faults shearing in the $\gamma'$ precipitate. It is well known that the isolated stacking faults are formed by shearing of the $\gamma'$ phase by $a/3<112>$ dislocations. In the current research, a great number of transition elements, such as Co and W are added to alloy 11, which reduces the stacking fault energy (SFE) of $\gamma$ matrix to some extent. According to Ref. [34], a reduction of SFE in $\gamma$ phase would facilitate the formation of continuous stacking fault. Therefore, both isolated stacking faults and continuous stacking faults were observed under the same crept alloy 11 samples.

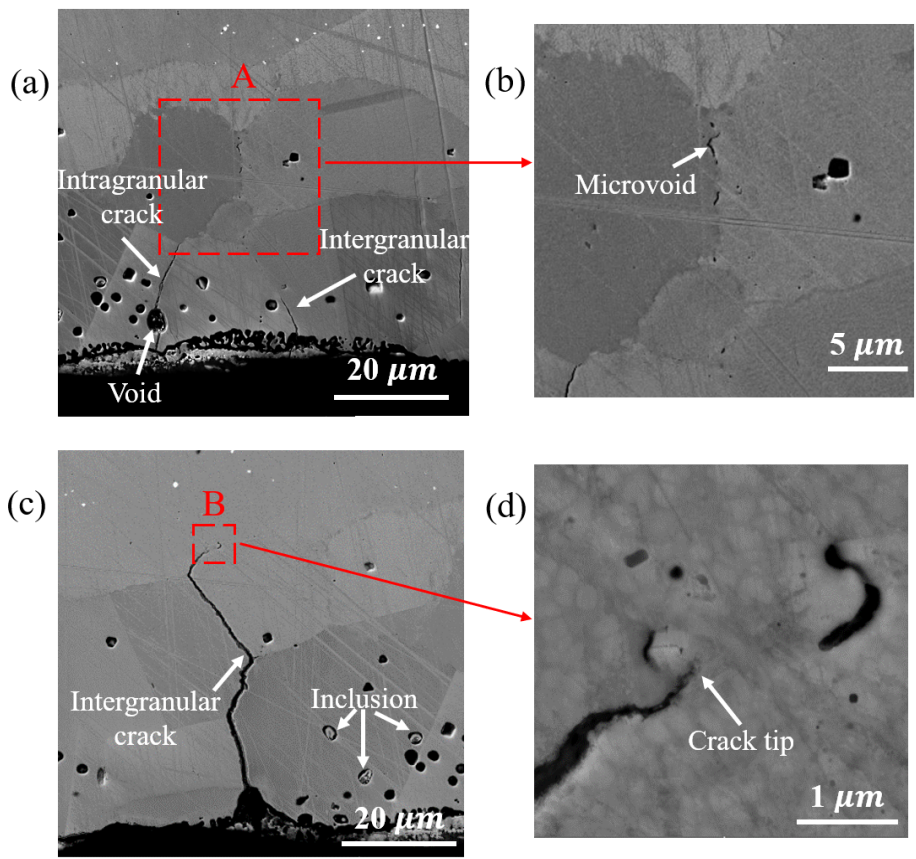

**Figure 6**. The microstructures and SEM images showing cracks, inclusions and microvoids in Sample B crept at 750°C/400MPa ((a) and (b)), and Sample C crept at 700°C/800MPa ((c) and (d)).

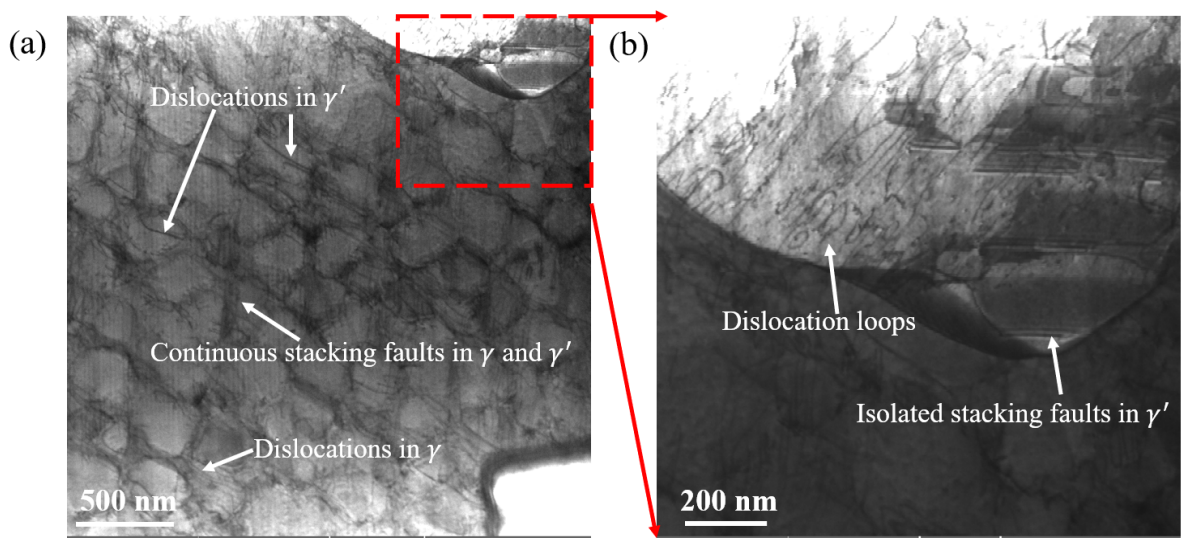

**Figure 7**. The microstructures and STEM images show the creep deformation mechanisms for alloy 11 after creeping at 700°C/800MPa for 0.26 hours.

## 3.2 Lattice misfit

For nickel-base superalloy, one important microstructural property is the $\gamma/\gamma'$ lattice misfit, $\delta$, which is defined as the relative difference of the lattice parameters in $\gamma$ and $\gamma'$ phase:

$$\delta = \frac{2(A_{\gamma'} - A_{\gamma})}{A_{\gamma'} + A_{\gamma}} \tag{1}$$

Where $A_{\gamma'}$ and $A_{\gamma}$ are the lattice parameters of the $\gamma'$ and $\gamma$ phase, respectively. All the misfits calculated in this research are constrained misfits and are directly derived from the X-ray diffraction data. In the reference specimen, the constrained misfit arises from the stress-free states with coherent $\gamma/\gamma'$ interfaces. In the pre-crept samples, the misfit represents a combination of stress-free constrained misfit and the residual strain contributions from the interfacial dislocation network.

Figure 8 (a) reports the contour map of lattice misfits in the three samples. Figure 8 (b) and (c) show the averaged lattice misfit in the horizontal and vertical direction, respectively. The lattice misfit in the reference sample is relatively uniform and ranges around -0.0031, while the lattice misfit in the two pre-crept samples shows a distinctly different spatial distribution. A much larger magnitude of negative misfit is observed in the middle region of crept Sample B and C. This particular distribution of misfit is due to the parabolic-shaped temperature distribution profile. Creep deformation is dependent on the applied stress and the temperature. Here the applied stress was constant at each cross-section that is perpendicular to the loading direction in the gauge length, while the temperature was non-uniform along the gauge length as discussed in Appendix A. Consequently, during the creep tests using the ETMT system, the sample centre had the highest temperature and is inclined to accumulate more deformation. In other words, creep deformation is localized when the sample is deformed using the ETMT system, and the predominant deformation occurs in the sample centre. The above discussion reveals that there might exist a threshold temperature (see Figure B in Appendix A) in ETMT system above which more creep deformation accumulates or the dominant creep deformation mechanism changes. The differences between the minimum and maximum misfit are 1.6%, 9.3% and 37.1% for Sample A, B and C, respectively. The considerable spatial variation in terms of misfit suggests that it is crucial to perform measurements at different sample locations with space intervals instead of a single measurement point when investigating lattice-related parameters, such as strains and stresses. Since Sample C has the highest difference in spatial misfit and also the largest creep rate, it is then reasonable to assume that more localized deformation is expected to accumulate in samples with a larger creep rate.

The evolution of lattice mismatch is time-dependent, i.e. Its value varies at different stages of creep. The result in the current study is comparable to an ex-situ neutron diffraction measurement of single crystal superalloy DD10 after creep [35] and an in-situ XRD measurement of superalloy AM1 during creep deformation [36], where the negative misfit ranges between -0.2% to -0.48% at different creep stages. The agreement suggests that the lattice misfit in superalloy developed during creep deformation is mostly retained after removing the external stress and thermal loading. In addition, the lattice misfit evaluated in pre-crept samples is also dependent on the magnitude of applied stress, as found in this study and other work in the literature [37].

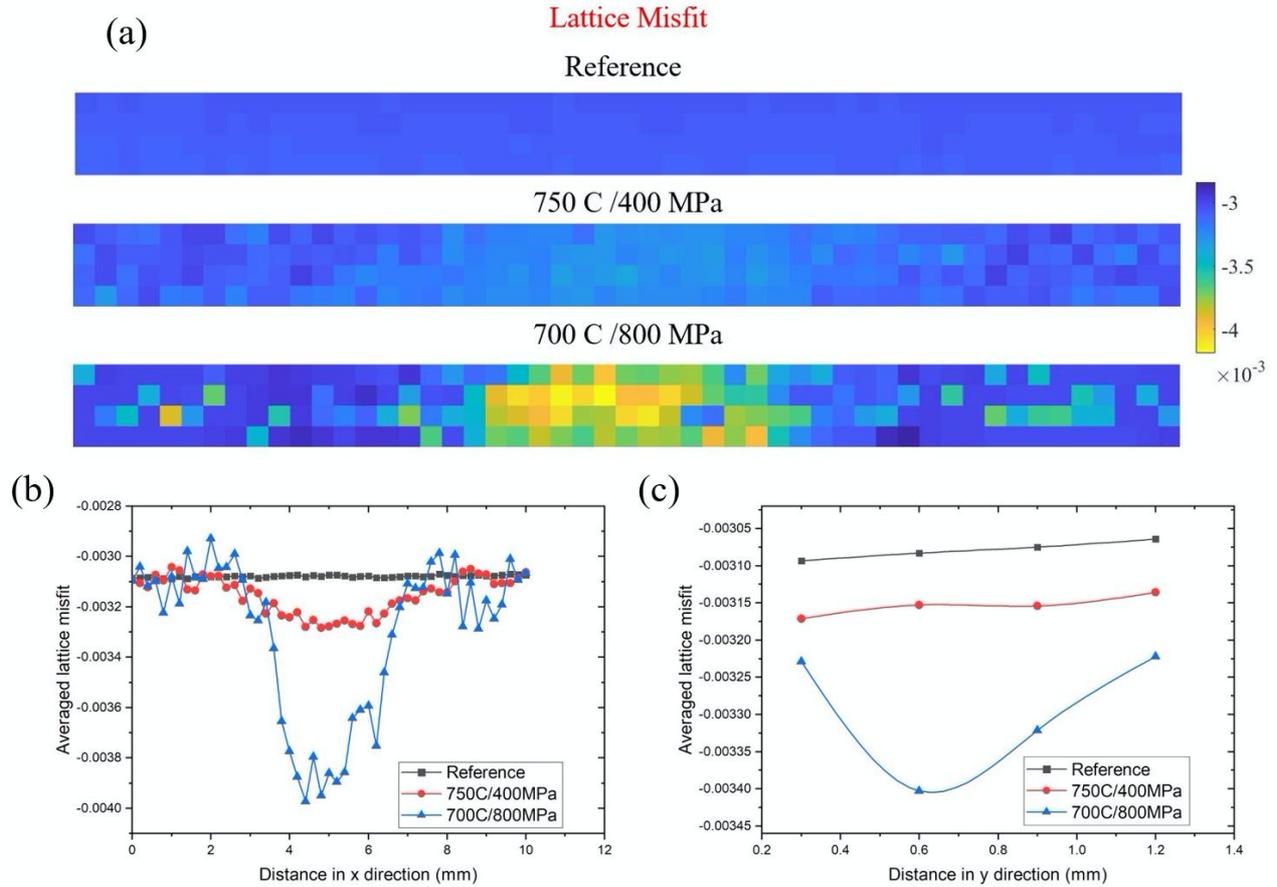

**Figure 8**. (a) The contour map of lattice misfits at different locations in the three samples. (b-c) The averaged lattice misfit along the x and y direction in the samples.

**3.3 Phase fraction**

Figure 9 (a) shows the contour map of the phase fraction of $\gamma$ in the three samples, and Figure 9 (b), and (c) illustrate the averaged $\gamma$ phase fraction in the horizontal and vertical directions, respectively. Analogous to lattice misfit, the phase fraction in the reference sample is relatively uniform and ranges around 64%, while the phase fraction of $\gamma$ in two pre-crept samples shows a slightly different spatial distribution and varies from 57% to 67%. The averaged $\gamma$ phase fractions in Sample A, B and C are 64%, 62% and 61%, respectively. The $\gamma'$ solvus temperature of this alloy is ~ 1100°C, which is much higher than the thermal condition in the creep test (750°C for Sample B, and 700°C for Sample C). The slight decrease in the $\gamma$ phase fraction after creep deformation indicates that there is a redistribution of alloying elements within the microstructure during creep even when the creep temperature is lower than the solvus temperature. Compared with Sample B, Sample C had a lower temperature, a much higher stress level and the lowest $\gamma$ phase fraction. This finding implies that element diffusion is a thermal process and external stress is likely to accelerate the evolution of $\gamma$ phase fraction.

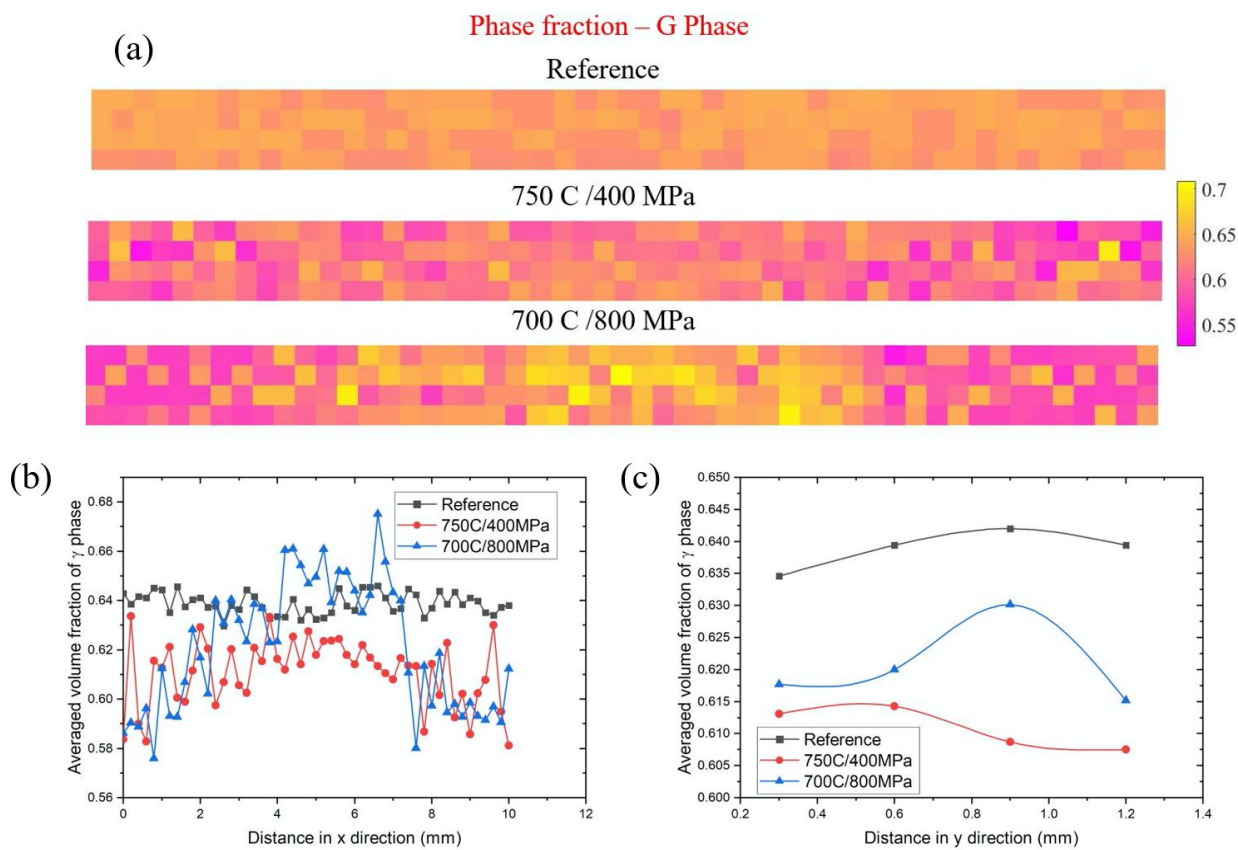

**Figure 9**. (a) The contour map of phase fraction of $\gamma$ in the three samples. (b-c) The averaged volume fraction of $\gamma$ phase along the x and y direction.

## 3.4 Microstrain

It is well known that microstrains could affect the broadness of the diffraction reflections and hence they were evaluated using the Scherer equation and strain function as implemented in GSAS-II. The microstrain in polycrystalline alloys reflects the lattice deformation and dislocation configuration within the alloys. Many publications suggest that the microstrain evaluated from XRD can be directly related to dislocation density in the crystalline materials [38-40]. Figure 10 and 11 illustrate the microstrain in $\gamma$ and $\gamma'$ phase in the three samples, respectively. The averaged microstrain in $\gamma$ phase in Sample A, B and C is 1465, 1598 and 2031, respectively, while the averaged microstrain in $\gamma'$ phase is 590, 1195 and 1360 in the samples, respectively. Here we demonstrate that the microstrain in $\gamma$ phase is much larger than the microstrain at the same location in $\gamma'$ phase in all three samples, which means that the dislocation density in $\gamma$ phase is much higher than that in $\gamma'$ phase for both reference and pre-crept samples.

It is clear from Figure 10(c) and Figure 11(c) that the microstrain for both $\gamma$ and $\gamma'$ phase in Sample C with the largest creep rate is higher than that in Sample B and A. As can be seen from Figure 10(b) and Figure 11(b), the centre of Sample C was observed to have the highest value of microstrain compared to the other locations and the other samples. More dislocation networks were developed in both $\gamma$ matrix and $\gamma'$ precipitate in the centre of Sample C, which might indicate the main creep deformation mechanism in Sample B and Sample C are different. In a high-temperature/low-stress regime like the loading condition in Sample B, creep deformation is dominantly caused by dislocation loops or climbs moving in the $\gamma$ matrix and forming interfacial dislocation networks at the $\gamma/\gamma'$ interfaces. This mechanism is active until the back stress of the dislocations accumulated in the networks makes it impossible for new dislocations to enter the $\gamma$ matrix channel. In low-temperature/high-stress regime like Sample C, the favourable deformation mechanism is the shearing of the microstructure (both $\gamma$ matrix and $\gamma'$ precipitate) by stacking faults. It is suggested that the dominant creep deformation mechanism in Sample C is the stacking faults cutting through the $\gamma$ and $\gamma'$ phase (as illustrated in Figure (7)), resulting in a higher dislocation density and microstrain in both phases as observed in this study.

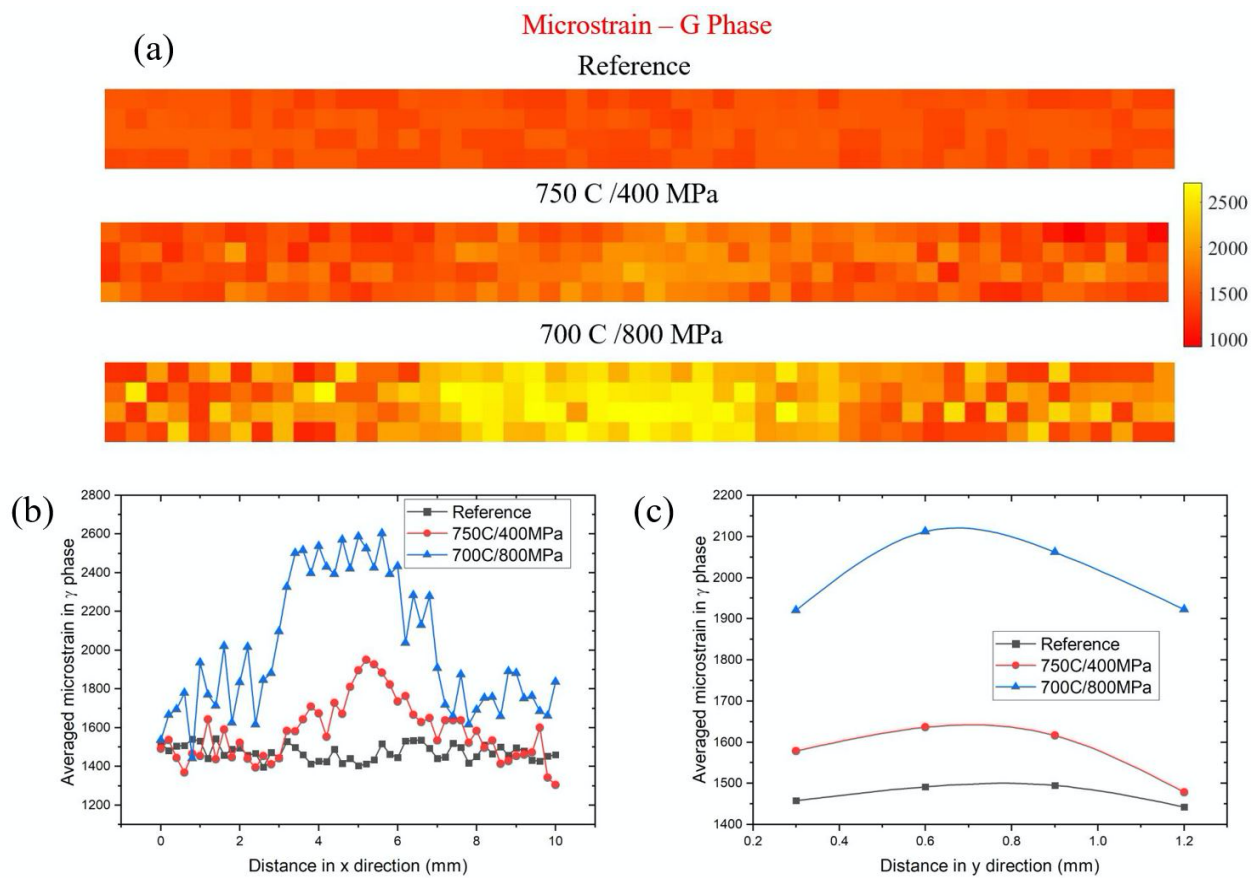

**Figure 10**. (a) The contour map of microstrain in $\gamma$ phase in the three samples. (b-c) The averaged microstrain in $\gamma$ phase along the x and y direction.

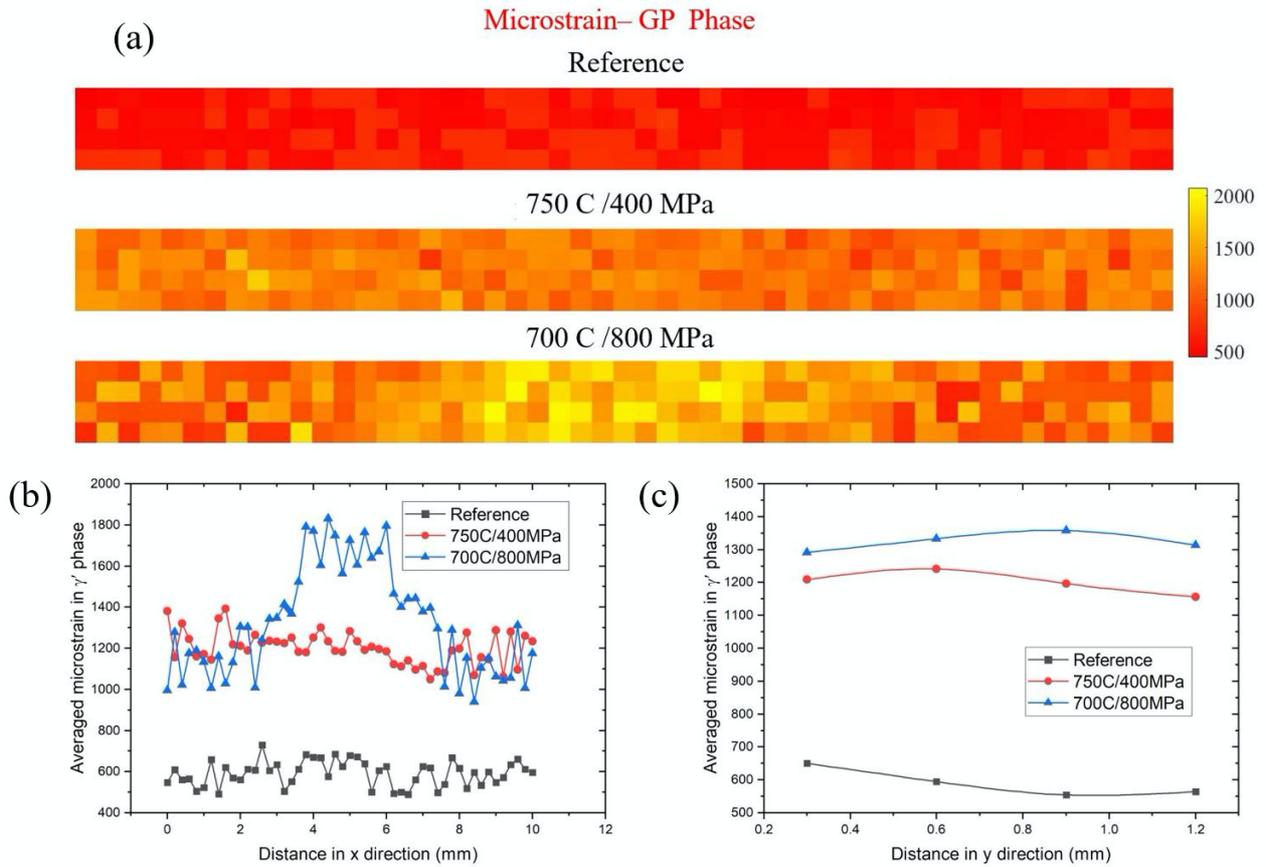

**Figure 11**. (a) The contour map of microstrain in $\gamma'$ phase in the three samples. (b-c) The averaged microstrain in $\gamma'$ phase along the x and y direction.

### 3.5 Lattice parameter distributions

### 3.5.1 Horizontal direction – lattice parameter

A quantitative determination of lattice parameters can be calculated from the relative change in X-ray diffraction peak position, which provides important information not only on the compositional change of the matrix and precipitate phase but also on the developed strain field. As creep is a thermomechanical process, the changes in lattice parameters can be caused by both the occurrence of compositional changes in the phases and the internal stress/strain field. It is then inaccurate to evaluate internal stress from the changes in lattice parameters, as reported in the literature [35]. In the current study, the XRD diffraction patterns were caked and integrated along a specified azimuth angle range to evaluate the lattice parameters in both $\gamma$ and $\gamma'$ in horizontal and vertical directions.

Figure 12 and 13 illustrate the horizontal lattice parameters in $\gamma$ and $\gamma'$ phase in the three samples, respectively. The lattice parameters in both phases for the reference sample are relatively constant and do not vary with different locations. The averaged horizontal lattice parameters in $\gamma$ phase in Sample A, B and C are 3.5981, 3.5971 and 3.5993, respectively, and the averaged horizontal lattice parameters in $\gamma'$ phase are 3.5864, 3.5862 and 3.5882 in the samples, respectively. An apparent finding is that the lattice parameters of $\gamma$ phase are all slightly larger than those of $\gamma'$ phase, which provides a negative value of lattice misfit. In the sample centre, the lattice parameter in the $\gamma$ phase in crept Sample C is larger than that in the reference sample, while the $\gamma$ lattice parameter in crept Sample B is smaller compared to the reference sample. Significant deviations were observed in the $\gamma'$ precipitate in the centre position of Sample C compared to the reference sample, while no apparent deviation was detected in the same region of Sample B, indicating the dominant creep deformation mechanism is different in this region for the two samples. The lattice parameter in Sample C varies significantly at different locations, this can be explained by the fact that the internal stresses and strains are orientation-dependent after creep loading. The orientation dependence will become more prominent when the grain is large and elongated (only a few grains in the region of interest) as for Sample C.

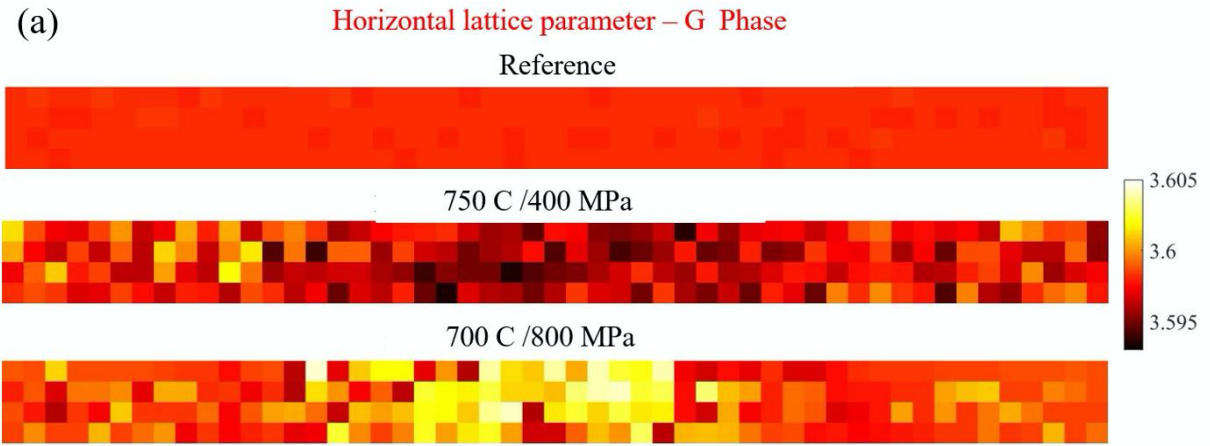
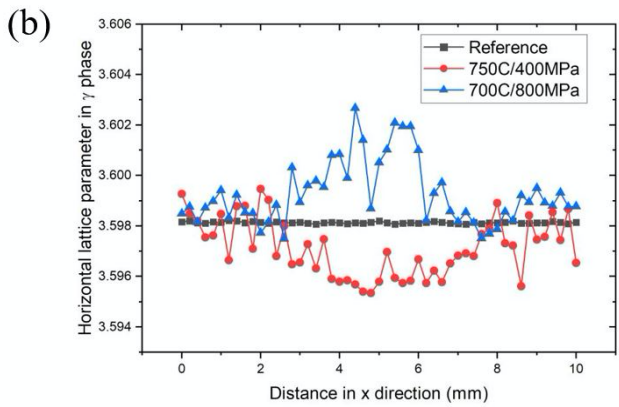
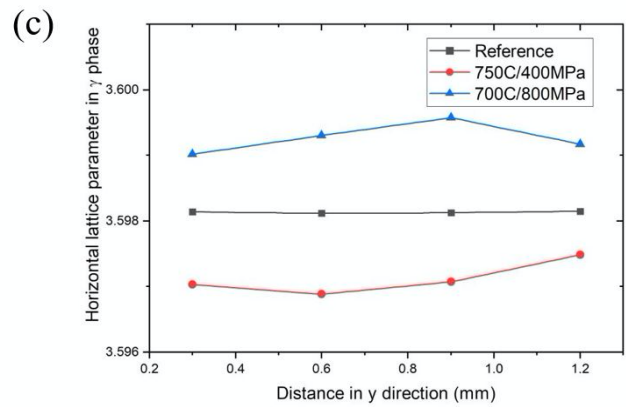

**Figure 12.** (a) The contour map of horizontal lattice parameter in $\gamma$ phase in the three samples. (b-c) The averaged horizontal lattice parameter in $\gamma$ phase along the x and y direction.

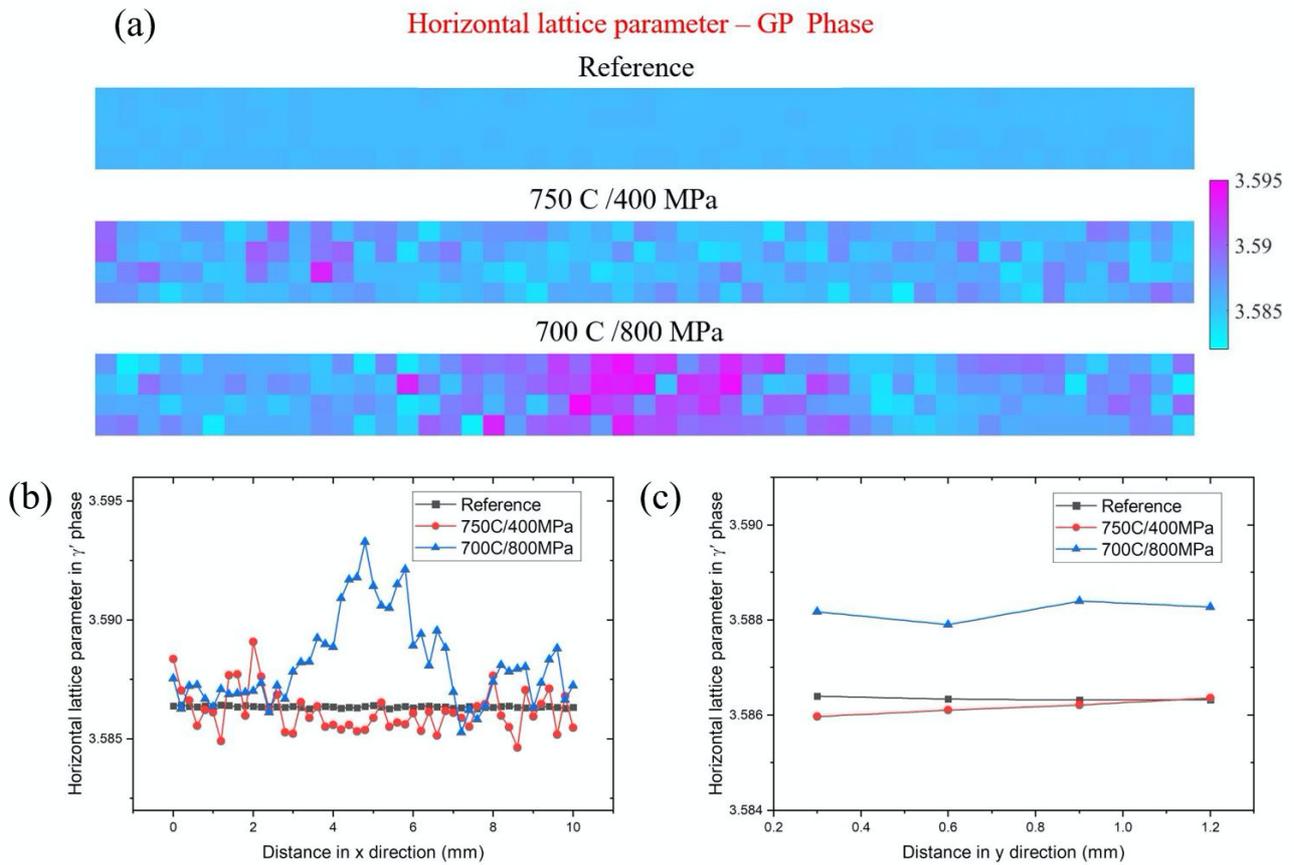

**Figure 13**. (a) The contour map of horizontal lattice parameter in $\gamma'$ phase in the three samples. (b-c) The averaged horizontal lattice parameter in $\gamma'$ phase along the x and y direction.

### 3.5.2 Vertical direction – lattice parameter

Figure 14 and 15 illustrate the vertical direction lattice parameter in $\gamma$ and $\gamma'$ phase in the three samples, respectively. There exists some extent of lattice parameter fluctuation in both phases for Sample B. A dramatic increase in lattice parameters of both phases is developed in the centre of Sample C. There is a prominent spatial variation of the lattice parameters in Sample C, suggesting that it is crucial to perform mapping measurement instead of a single measurement point when investigating the lattice-related parameters.

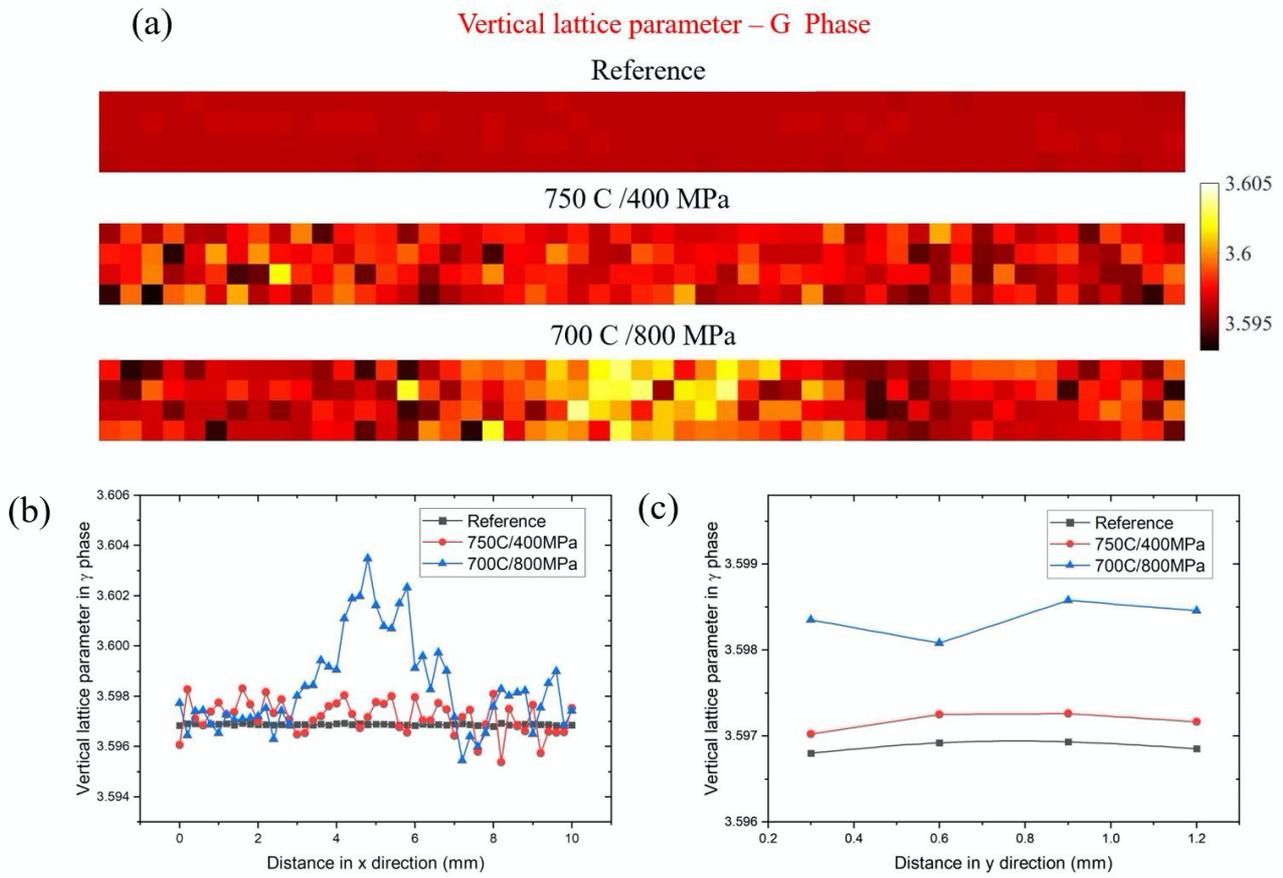

**Figure 14**. (a) The contour map of vertical lattice parameter in $\gamma$ phase in the three samples. (b-c) The averaged vertical lattice parameter in $\gamma$ phase along the x and y direction.

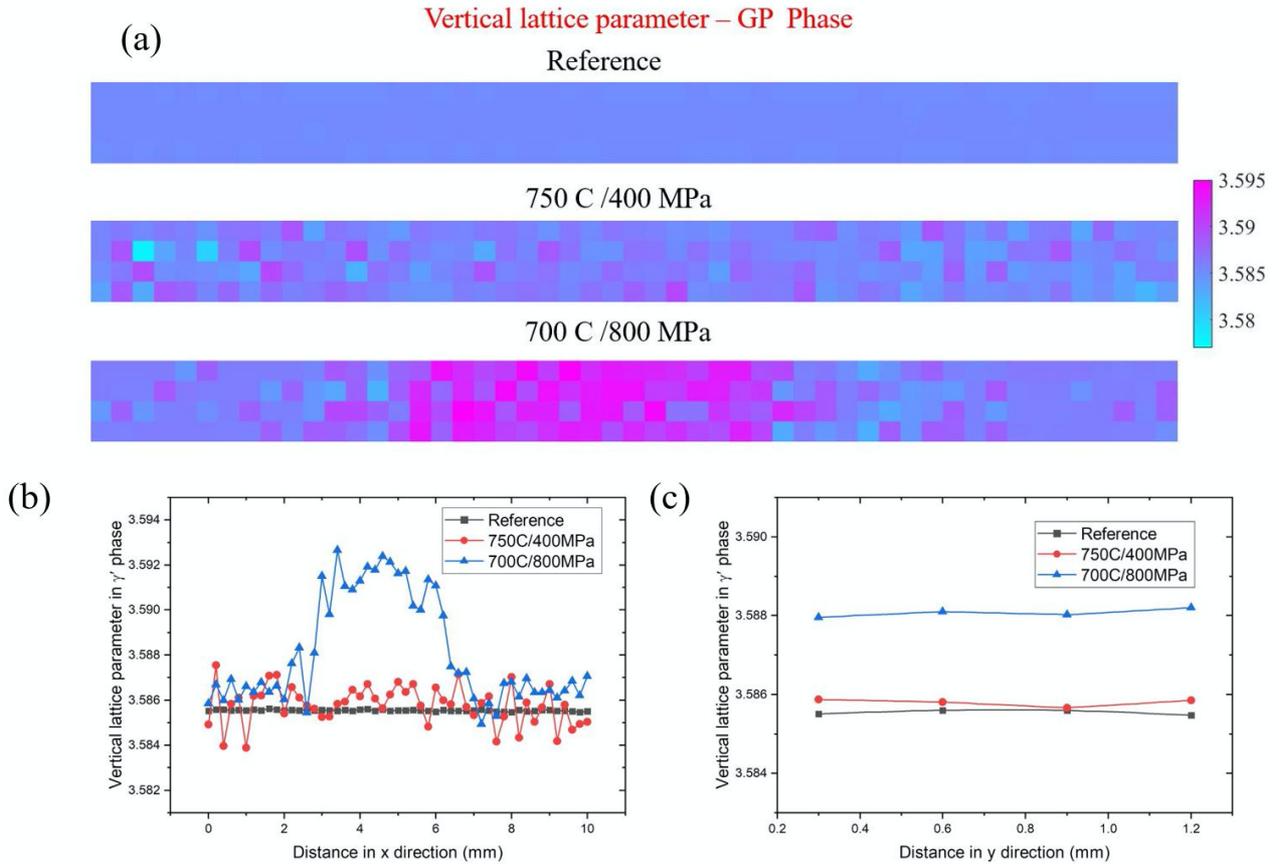

**Figure 15**. (a) The contour map of vertical lattice parameter in $\gamma'$ phase in the three samples. (b-c) The averaged vertical lattice parameter in $\gamma'$ phase along the x and y direction.

## 3.6 The statistical distribution of lattice parameter variation

In polycrystalline materials such as nickel-base superalloys, the elastic and plastic deformation are highly anisotropic, giving rise to strain spatial inhomogeneity that emerges within and between grains. The inhomogeneous nature of strain distribution has led to the concept of three types of strains that describe the spatial variation over different length scales: macroscale, mesoscale and microscale. On the macroscale, the whole sample can be regarded as a homogeneous medium with averaged material properties such as Young's modulus and yield stress. On the atomic scale (microscale), deformation is characterized by dislocation motion, dislocation-dislocation and dislocation-precipitate interactions. In the current study, we are interested in the deformation behaviour at the mesoscale where the controlling microstructural features are grain size, morphology and orientation, dislocation density, phase properties, and their spatial distribution. The investigation of the distribution of lattice-parameter change at the mesoscale can effectively provide further insights into the effect of mesoscopic microstructure on mechanical properties. The lattice-parameter change $\varepsilon$ can be defined as follows:

$$\varepsilon = \frac{a - a_0}{a_0} \qquad (2)$$

Where $a$ and $a_0$ denote the lattice parameters at the loaded and stress-free state, respectively. Figure 16 illustrates the statistical distribution of horizontal lattice-parameter change in $\gamma$ and $\gamma'$ phase in Sample B. It is interesting to find that the lattice-parameter changes for both phases in the loading direction follow a normal distribution. The same type of statistical distribution of lattice-parameter change is observed in Sample C as shown in Figure 17. In this experiment, the sample thickness is 1.2mm and the X-ray beam size is $200\,\mu m \times 200\,\mu m$, leading to an illumination volume of $200\mu m \times 200\mu m \times 1200\mu m$ for a single diffraction pattern. Each measurement point provides averaged values of material properties and mechanical response within the illumination volume. Each illumination volume contains dozens to thousands of grains, and hence, it can be regarded as a representative volume element (RVE) for the material investigated in this study. The lattice-parameter change in the current study arises from a complex thermomechanical process that includes the effects of both strain field and compositional change. Our previously published simulation results suggest that the flow stresses obtained from multiple RVEs with nominally identical microstructures conform to a normal distribution [41,42]. The finding in this research further extends the applicability of our simulation results from only mechanical loading conditions to thermomechanical processes in polycrystalline materials.

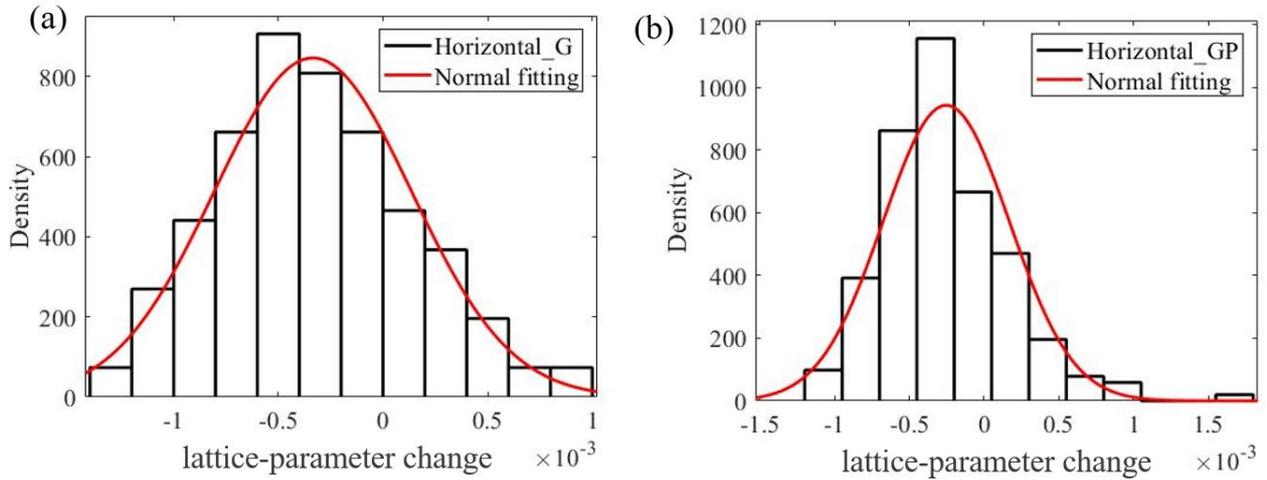

**Figure 16**. The statistical distribution of lattice-parameter changes in the loading direction within (a) $\gamma$ and (b) $\gamma'$ phase in Sample B crept at 750°C/400MPa.

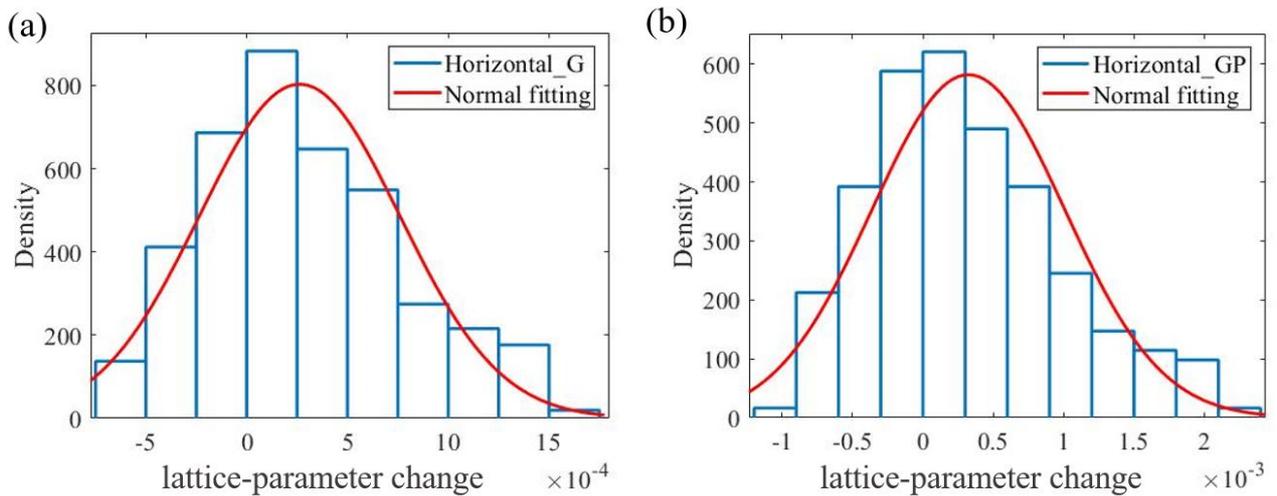

**Figure 17**. The statistical distribution of lattice-parameter changes in the loading direction within (a) $\gamma$ and (b) $\gamma'$ phase in Sample C crept at 700°C/800MPa.

# 4. Conclusion

The creep behaviours of a newly designed nickel-base superalloy at 750 °C/400MPa and 700 °C/800MPa have been systematically investigated using the scanning electron microscope, scanning transmission electron microscopy, electron backscatter diffraction and synchrotron X-ray diffraction. The experiment results from pre-crept samples were compared with the results for the reference sample. Material properties and mechanical responses such as lattice misfit, phase fraction, lattice parameter and microstrain in the reference and crept samples have been analysed in detail. The main conclusions of this paper can be listed as follows:

(1) The materials characterization of pre-crept samples reveals that creep loading conditions can introduce detwinning and recrystallization in nickel-base superalloy. A higher magnitude of external loading stress not only dramatically increases the creep rate but also accelerates the dynamic recrystallization process.

(2) Both intergranular and intragranular cracks were found in the pre-crept samples. Four types of creep deformation mechanisms, namely dislocations in the $\gamma$ matrix channel, shearing of $\gamma'$ precipitate by APBs, continuous and isolated stacking faults, were observed in the sample crept at high stress and low temperature.

(3) The highest values of lattice misfit and microstrain were observed at the centre of crept samples, which was caused by the parabolic-shaped temperature distribution profile developed across the width of the sample during creep test using ETMT system. The significant spatial variation in terms of misfit, microstrain and lattice parameters suggests that it is crucial to perform measurements at different sample locations instead of a single measurement point when investigating the lattice-related parameters.

(4) In both crept samples, the statistics of the lattice-parameter changes for $\gamma$ and $\gamma'$ phases in the loading direction were found to follow a normal distribution.

**Acknowledgements**

The authors would like to acknowledge the award of the synchrotron X-ray beam time for the experiments in Diamond Light Source (MG25467-3). AMK wishes to acknowledge EPSRC support under grant EP/V007785/1, as well as the teaching materials for paper A4 Energy Systems, part of the undergraduate course in Engineering Science, University of Oxford.

# Appendix A

# Evaluation of the temperature distribution during creep testing using ETMT

The temperature distribution within the sample during creep testing using ETMT is known to be non-uniform and requires evaluation for result interpretation.

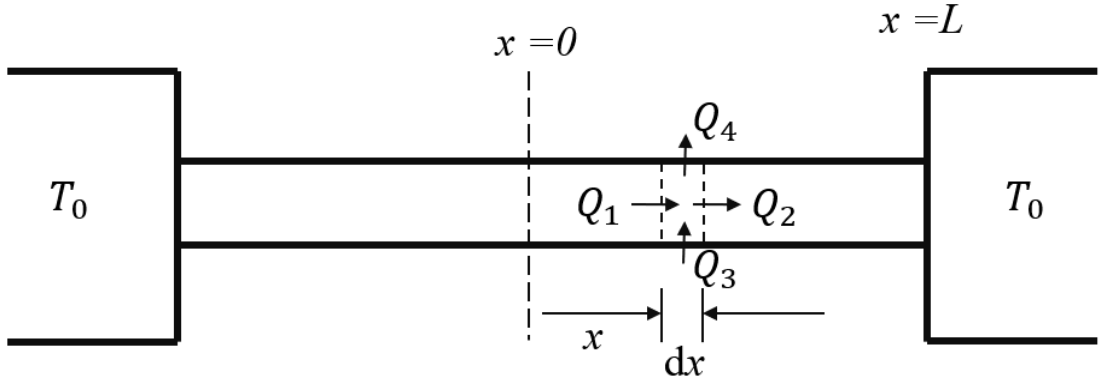

**Figure A**. The illustration of the heat transfer process in an ETMT sample.

Considering the 1D representation of the sample as a wire of diameter $2R$ and length $2L$ illustrated in Figure A, assume that the cooled grips are maintained at a fixed ambient temperature $T_0$. Let the sample thermal conductivity be denoted by $k$, and let Joule heating per unit time and volume be given by $G$. Assuming that radiative and convective heat loss can be represented by the heat transfer coefficient $h$, the heat balance equation can be written as

$$k\pi R^2 \frac{dT}{dx} - k\pi R^2 \left[\frac{dT}{dx} + \frac{d^2T}{dx^2}dx\right] + \pi R^2 \, dx \, G - 2\pi R \, dx(T-T_0)h = 0 \tag{3}$$

Introducing the temperature parameter $\theta = T - T_0$ the above equation is reduced to

$$\frac{d^2\theta}{dx^2} = \frac{G}{k} - \frac{2\theta h}{Rk} \tag{4}$$

Equation (4) has the solution

$$\theta = A \sinh mx + B \cosh mx + \frac{GR}{2h}, \text{ where } m = \sqrt{\frac{2h}{Rk}} \tag{5}$$

To complete the solution, we determine the coefficients as follows: at $x = 0, \frac{d\theta}{dx} = 0$, hence $A = 0$.

At $x = L, \theta = T - T_0 = 0,$ and hence $B = -\frac{GR}{2h}\frac{1}{\cosh mL}$.

Therefore, the temperature $T$ at location $x$ can be calculated as

$$T = T_0 + \frac{GR}{2h}\left[1 - \frac{\cosh mx}{\cosh mL}\right] \tag{6}$$

Differentiating both sides with respect to $x$ yields the following expression:

$$\frac{dT}{dx} = \frac{GR}{2h} m \frac{\sinh mx}{\cosh mL} \tag{7}$$

At $x = L$, $\left.\frac{dT}{dx}\right|_{x=L} = \frac{GR}{2h} m \tanh mL$.

The heat lost through one end is equal to

$$Q_{end} = k\pi R^2 \left.\frac{dT}{dx}\right|_{x=L} = \frac{k\pi R^3 Gm}{2h} \tanh mL \tag{8}$$

But $L = \sqrt{\frac{kR}{2h}}$, so $mL = 1$ and $\tanh mL = 0.7616$.

Hence,

$$Q_{end} = 0.7616\, \pi R^2 LG \tag{9}$$

The heat generated in half the wire is $\pi R^2 LG$, which means that 76% of the generated heat is lost through the ends. As illustrated in Figure B, the analytical solution discussed in Appendix A provides an excellent match to the experimentally measured temperature in the creep test using the ETMT system [43]. The fitting parameters are listed in Table 3.

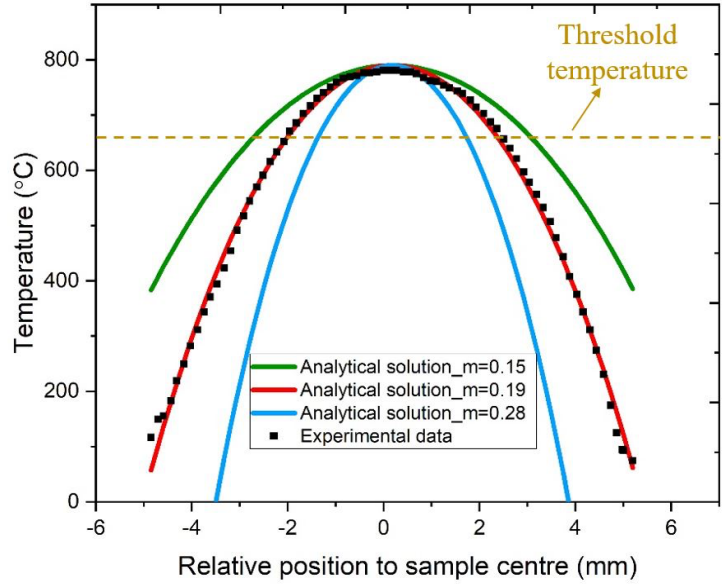

**Figure B.** The temperature profile along the sample obtained from experimental measurement during a creep test [43] and the best fit of the analytical solution is presented here.

Table 3. The fitting parameters for the analytical solution.

| Parameters | $T_0$ (K) | $L$(m) | $R$(m) | $m\ (m^{-1})$ | $k$ (W/(m·K)) | $h$ (W/($m^2$ ·K)) | $G$ (J/(s · $m^3$)) |
|---|---|---|---|---|---|---|---|
| Value | 293.15 | $1 \times 10^{-2}$ | $7.6 \times 10^{-4}$ | 0.19 | 9.8 | $1.3 \times 10^{-4}$ | 73.4 |

As illustrated in Figure B using multiple curves, the key parameter $m$ defines the steepness of the temperature profile variation in the middle of the sample. Parameter $m$ depends on the sample geometry, namely, its effective wire radius $R$, and the sample thermal properties: conductivity $k$ and the heat transfer coefficient between the sample and environment, $h$. Since parameter $m$ <u>does not</u> depend on the maximum temperature of the experiment, nor the heat generation parameter $G$, the temperature variation curves for different experiments using identical samples can be obtained from the best fit 'master' curve shown above by scaling the temperature range.